\documentclass[aps,prx,superscriptaddress,twocolumn,10pt]{revtex4-2}
\usepackage[titletoc,toc,title,page]{appendix}

\usepackage{csquotes}
\usepackage[italian,english]{babel}

\usepackage{microtype} 
\usepackage{bm} % provides more bold symbols

\usepackage{dsfont} % provides more \mathbb \smallskipymbols
\usepackage{amsmath,amssymb,amsthm,thmtools}
\usepackage{mathtools}
\usepackage{cases}
\usepackage{calc}
\usepackage{mathrsfs} % provides the nice \matchsrc font
\usepackage[normalem]{ulem} %provides \sout command for striking through text
\usepackage{nameref}
\usepackage[colorlinks=true]{hyperref}
\usepackage[nameinlink]{cleveref}
\crefname{appsec}{Appendix}{Appendices}
\crefname{box}{Box}{Box}
\hypersetup{
  colorlinks   = true, %Colours links instead of ugly boxes
  urlcolor     = green!80!black, %Colour for external hyperlinks
  linkcolor    = blue, %Colour of internal links 	q
  citecolor    = red!80!black %Colour of citations
}

\usepackage{physics} % provides Dirac notation and lots of other things

\usepackage{float} % provides the option H for includegraphics

\usepackage{graphicx}

\usepackage[usenames,dvipsnames,table]{xcolor}
\usepackage{easyReview}

\usepackage{tikz}
\usetikzlibrary{calc,shapes.geometric}

\usepackage{placeins}
\usepackage{multirow,tabularx,booktabs}
\usepackage[most]{tcolorbox}
\newtcbtheorem{tbox}{Box}{enhanced, float*=t, width=\textwidth, label type=box}{box}

\usepackage[printonlyused,withpage,nohyperlinks,smaller]{acronym}

\graphicspath{{./figures/}}

\newcommand{\RR}{\mathbb{R}}
\newcommand{\CC}{\mathbb{C}}
\newcommand{\NN}{\mathbb{N}}

\newcommand{\rmC}{\mathrm{C}}
\newcommand{\rmD}{\mathrm{D}}

\newcommand{\rmU}{\mathrm{U}}

\newcommand{\calE}{\mathcal{E}}

\newcommand{\calO}{\mathcal{O}}

\newcommand{\calH}{\mathcal{H}}

\newcommand{\calX}{\mathcal{X}}
\newcommand{\calY}{\mathcal{Y}}

\newcommand{\bs}[1]{\boldsymbol{#1}}
\newcommand{\on}[1]{\operatorname{#1}}
\newcommand{\parTitle}[1]{\noindent{\color{Mahogany}(\emph{#1})}}

\newcommand{\bstildemu}{\bs{\tilde{\mu}}}

\newcommand{\addLI}[1]{{#1}}

\DeclareMathOperator{\Herm}{Herm}

\renewcommand{\parTitle}[1]{}
\begin{document}
\title{Potential and limitations of quantum extreme learning machines}
\author{L. Innocenti}
\let\comma,
\affiliation{Universit\`a degli Studi di Palermo\comma{} Dipartimento di Fisica e Chimica - Emilio Segr\`e\comma{} via Archirafi 36\comma{} I-90123 Palermo\comma{} Italy}
\author{S. Lorenzo}
\affiliation{Universit\`a degli Studi di Palermo\comma{} Dipartimento di Fisica e Chimica - Emilio Segr\`e\comma{} via Archirafi 36\comma{} I-90123 Palermo\comma{} Italy}
\author{I. Palmisano}
\affiliation{Centre for Theoretical Atomic\comma{} Molecular\comma{} and Optical Physics\comma{} School of Mathematics and Physics\comma{} Queen's University Belfast\comma{} BT7 1NN\comma{} United Kingdom}
\author{A. Ferraro}
\affiliation{Centre for Theoretical Atomic\comma{} Molecular\comma{} and Optical Physics\comma{} School of Mathematics and Physics\comma{} Queen's University Belfast\comma{} BT7 1NN\comma{} United Kingdom}
\affiliation{Quantum Technology Lab\comma{} Dipartimento di Fisica Aldo Pontremoli\comma{} Universit\`a degli Studi di Milano\comma{} I-20133 Milano\comma{} Italy}
\author{M. Paternostro}
\affiliation{Centre for Theoretical Atomic\comma{} Molecular\comma{} and Optical Physics\comma{} School of Mathematics and Physics\comma{} Queen's University Belfast\comma{} BT7 1NN\comma{} United Kingdom}
\author{G. M. Palma}
\affiliation{Universit\`a degli Studi di Palermo\comma{} Dipartimento di Fisica e Chimica - Emilio Segr\`e\comma{} via Archirafi 36\comma{} I-90123 Palermo\comma{} Italy}
\affiliation{NEST\comma{} Istituto Nanoscienze-CNR\comma{} Piazza S. Silvestro 12\comma{} 56127 Pisa\comma{} Italy}

 \date{\today}

\begin{abstract}
    Quantum extreme learning machines (QELMs) aim to efficiently post-process the outcome of fixed --- generally uncalibrated --- quantum devices to solve tasks such as the estimation of the properties of quantum states.
    The characterisation of their potential and limitations, which is currently lacking, will enable the full deployment of such approaches to problems of system identification, device performance optimization, and state or process reconstruction.
    We present a framework to model QELMs, showing that they can be concisely described via single effective measurements, and provide an explicit characterisation of the information exactly retrievable with such protocols.
    We furthermore find a close analogy between the training process of QELMs and that of reconstructing the effective measurement characterising the given device.
    Our analysis paves the way to a more thorough understanding of the capabilities and limitations of QELMs, and has the potential to become a powerful measurement paradigm for quantum state estimation that is more resilient to noise and imperfections.
\end{abstract}
\maketitle

\section{Introduction}

\parTitle{Classical RCs and ELMs}
Extreme learning machines (ELMs) \cite{huang2004extreme,huang2011extreme,wang2021review}
and Reservoir computers (RC) \cite{lukovsevivcius2009reservoir,lukosevicius2012practical,angelatos2021reservoir,jaeger2001echo,jaeger2004harnessing,van2017advances} 
are computational paradigms that leverage fixed, nonlinear dynamics to efficiently extract information from a given dataset.
In the classical context, these schemes rely on evolving input data through some nonlinear mapping --- typically recurrent neural networks with fixed weights --- which augment the dimensionality of the data, easing the extraction of the properties of interest.
The main discriminator between RCs and ELMs is whether the reservoir being used can deploy an internal memory.
% whether the map applied to the input vectors holds internal memory.
More precisely, RCs hold memory of the inputs seen at previous iterations, making them suitable for {temporal data processing}~\cite{lukosevicius2012practical}.
% In this context, a commonly used approach are the so-called \textit{echo state networks} (ESNs)~\cite{jaeger2001TheechoST,martinezpena2021dynamical}.
ELMs instead use memoryless reservoirs. Although this makes the training of ELMs easier, it also makes them unsuitable for temporal data processing.

\parTitle{How do RCs work\\
Broadly speaking, RCs and ELMs both comprise a \textit{reservoir} and a \textit{readout stage}.
The reservoir implements a mapping on the input data, and is characterised by fixed, often randomly pre-chosen, parameters. The readout stage is usually a single neural network layer, whose parameters are trained for example via linear regression.
% First, a recurrent neural network (RNN) with random, untrained weights is used to \textit{scramble} the input data; in this phase the data is mapped nonlinearly into a larger space.
% Then a \textit{readout phase} follows, in which the neurons of the reservoir are connected to an output layer.
The neurons of this output layer encode the output of the RC. The trained parameters are the weights of the connections between reservoir and output layer. Two aspects of RC schemes are worth remarking: the nonlinear ``scrambling'' dynamics offered by the fixed-weights reservoir, and the memory effects brought about by the natural flow of data through a reservoir, which can induce correlations between the outputs at different iterations.
Such memory is the main difference between ELMs and RCs: in the former, the reservoir has no memory capacity, and there is thus no correlation between outputs at different iterations.
% (more precisely put, the output at each iteration depends solely on the input at the same iteration).
}

\parTitle{Quantum reservoir computing, overview}
Quantum counterparts to ELMs and RCs --- which we will refer to as QELMs and QRCs, respectively --- have recently attracted significant interest due to their potential to process quantum information~\cite{fujii2017harnessing,%
ghosh2019quantum,ghosh2021realising,% ELMs
martinezpena2020information,kutvonen2020optimizing,tran2020higherorder,martinezpena2021dynamical,krisnanda2021creating,%
rafayelyan2020largescale,rafayelyan2020largescale,%optical
nokkala2021gaussian,%continuous variables
nakajima2019boosting,mujal2022quantum,mujal2023time,martinez2023quantum}.
Reviews of the state of the art in this context can be found in Refs~\cite{tanaka2019recent,fujii2020quantum,mujal2021opportunities}, while a study of QRC schemes for the implementation of nonlinear input-output maps with memory on NISQ devices has recently been presented~\cite{chen2020temporal}.

\parTitle{In this paper...}
\parTitle{What the literature is missing}
To date, and to the best of our knowledge, a general characterisation of the class of tasks that can be accomplished through QELM-like schemes for the classification, processing, or extraction of information encoded in quantum states is lacking. This significantly limits the systematic deployment of such approaches to the issues of quantum-system and quantum-state characterization or validation, which are crucial steps to perform towards the upscaling of quantum technologies and the achievement of the fault-tolerant quantum information processing paradigm.

In this paper, we show that the problem of reconstructing features of a quantum state via an ELM-like setup can be viewed as a linear regression task on the measurement probabilities produced by a suitable positive operator valued measurement (POVM) \cite{ghosh2020reconstructing,Nakajimatemporal2021}.
The key observation is that %quantum operations are necessarily linear --- more precisely, 
the probability distribution corresponding to an arbitrary measurement of a quantum state is linear in the input density matrix~\cite{watrous2018theory}.
This is a fundamental departure from \textit{classical} ELMs: whereas in the latter case the reservoir is an intrinsically {nonlinear} operation, the same cannot be said about a quantum reservoir. The latter can always be modelled as a map that linearly processes  the input density matrix.
In turn, this allows us to identify crucial constraints on the properties that QELM setups can be trained to retrieve.
% , which are bound to be linear in the input states.
While the learning of classical input information that is nonlinearly encoded in the states~\cite{nokkala2021gaussian,govia2021nonlinear} is certainly not precluded, our study clarifies how the only possible source of nonlinearity must come from the encoding itself rather than the reservoir dynamics.

We then show that the intrinsic uncertainty arising from the sampling noise on estimated measurement probabilities dramatically affects the performances of any property-reconstruction protocol based on 
QELMs. This pinpoints a significant fundamental constraint  -- of strong experimental relevance -- to the performance of such schemes. 
The number of measurement outcomes is also shown to play an important role, affecting the well-conditioning of the associated regression problem, and thus the numerical stability of any estimate.
\addLI{More generally, we show that the efficiency of QELMs is directly tied to the effective POVM summarizing both evolution and measurement. This puts the spotlight on the properties of this effective POVM, and on how these are the ones directly affecting performances.}

% In this regard, we discuss how suitable choices of effective POVMs can be used to improve overall performances. 

By addressing fundamental features of significant practical repercussions, our study allows to shape the contours of the class of tasks that can be successfully tackled through such novel architectures for quantum information processing, and contributes to the investigation of property-reconstruction protocols, assisted by artificial intelligence, which is raising growing  attention from thew quantum-technology community.

The remainder of this paper is organized as follows. In Sec.~\ref{sec:QELM_single_injection} we set the context of QELMs and provide both the main formal results of our analysis, and a reconstruction method, whose efficiency we briefly discuss. In Secs.~\ref{sec:single-inj-examples} and \ref{sec:QELM_multiple_injections} we address the cases of single and multiple injections of the input state, assessing the capacity of QELMs to reconstruct a given target observable, providing analytical and numerical results when considering both linear and nonlinear functionals of the input density matrix. Finally, in Sec.~\ref{conc} we draw our conclusions. 

\section{Description of the general QELM approach}
\label{sec:QELM_single_injection}

\parTitle{Section overview}
We here review the basic features of classical ELMs and QELMs, present a general way to model QELMs, and characterise their predictive power in various scenarios.

\subsection{Introduction and notation}

\begin{figure}[b!]
    \centering
    {\bf (a)}\\
        \includegraphics[width=\linewidth]{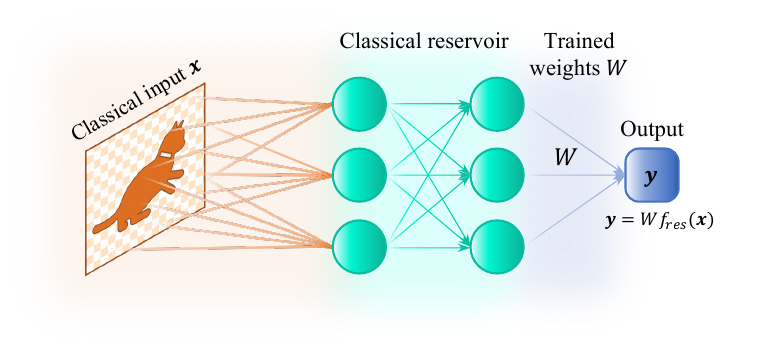}
    {\bf (b)}\\
        \includegraphics[width=\linewidth]{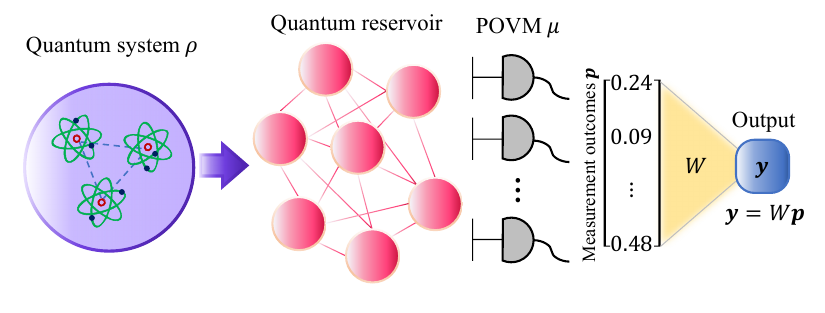}
    \caption{Schemes of principle of 
        {\bf (a)}: Classical ELM setup;
        %Middle: QELM scheme in metrological context. 
        {\bf (b)}: QELM setup for classical information processing.
    }
    \label{fig:classical_vs_quantum_RC}
\end{figure}

\parTitle{Classical ELM}
An ELM~\cite{huang2004extreme,wang2021review} is a supervised machine learning protocol which, given a training dataset $\{(\bs x_k^{\rm tr},\bs y_k^{\rm tr})\}_{k=1}^{M_{\rm tr}}\subset\RR^n\times\RR^m$, is tasked with finding a \textit{target function} $f_{\rm target}:\RR^n\to \RR^m$ such that, for each $k$, $f_{\rm target}(\bs x_k^{\rm test})\simeq \bs y_k^{\rm test}$ with a sufficiently good approximation for previously unseen datapoints $\{(\bs x_k^{\rm test},\bs y_k^{\rm test})\}_{k=1}^{M_{\rm test}}$.
As most machine learning algorithms, ELMs are characterised by their \textit{model}, that is, the way the input-output functional relation is parametrised. For ELMs, the model is a function of the form $\bs x\mapsto W f(\bs x)$ with $f$ a fixed --- generally nonlinear --- function implementing the reservoir dynamics, and $W$ a linear mapping applied to the output of $f$.
The function $f$ is \textit{not trained}, but rather fixed beforehand, and can for example be implemented as a neural network with fixed random weights.
The training algorithm optimises the parameters defining $W$ in order to minimise some distance function --- often the standard Euclidean distance --- between $W f(\bs x_k^{\rm tr})$ and $\bs y_k^{\rm tr}$.
As a classical example, one can think of a supervised learning task where $\bs x_k^{\rm tr}$ are images representing handwritten digits, and $\bs y_k^{\rm tr}$ the digits the images represent. In this example, $n$ would be the number of pixels in each image, and the goal of the algorithm would be to use the training dataset of labelled images $\{(\bs x_k^{\rm tr},\bs y_k^{\rm tr})\}_{k=1}^{M_{\rm tr}}$ to find the $W$ such that, for all new images $\bs x_k$, $W f(\bs x_k)$ is the correct digit drawn in $\bs x_k$.

\parTitle{QELMs}
The standard way to \textit{quantise} ELMs is to replace the map $f$ with some quantum dynamics followed by a measurement. To maintain full generality, we consider a completely positive trace-preserving (CPTP) quantum map $\Lambda$ --  which we refer to as a {\it quantum channel}  --- followed by a POVM  $\{\mu_b : b\in\Sigma\}$, where $\Sigma$ is the set of possible measurement outcomes~\cite{watrous2018theory}.
In the context of QELMs, the training dataset has the form $\{(\rho_k^{\rm tr}, \bs y_k^{\rm tr})\}_{k=1}^{M_{\rm tr}}$, with $\rho_k^{\rm tr}$ an input state and $\bs y_k^{\rm tr}$ the output vector that the QELM should associate to $\rho_k^{\rm tr}$.
More precisely, the goal of the training is to find a linear operation $W$ such that
\begin{equation}
    \sum_{b\in \Sigma} W_{ab}
    \Tr(\mu_b \Lambda(\rho_k^{\rm tr}))
    \simeq (\bs{y}_k^{\rm tr})_a,
\end{equation}
with $a=1\dots m$ and $k=1\dots M_{\rm tr}$, and with $W_{ab}$ the matrix elements of $W$.
It is also possible to use QELMs as a way to process classical information exploiting complex quantum dynamics.
In this case, the training dataset should be considered as a set of the form $\{(\bs s_k^{\rm tr}, \bs y_k^{\rm tr})\}_k$, in direct analogy with the classical case, where now $\bs s_k$ are classical vectors suitably encoded in the input quantum states $%\rho=
\rho_{\bs s}$. The difference with the classical setup, in this case, is entirely in the specific form of the function mapping inputs to outputs.
The capabilities of QELM/QRCs to process classical data depends crucially on the nonlinearity of the encoding $\bs s\mapsto \rho_{\bs s}$, as discussed in Refs.~\cite{nokkala2021gaussian,govia2021nonlinear} [cf.~\cref{fig:classical_vs_quantum_RC} for a schematic overview of the distinction between ELM and QELM protocols].
We will focus here on the former point of view to derive results that are independent of the specific forms of classical encodings $\rho_{\bs s}$, and useful when the goal is to probe property of the input states.

The ``classical reservoir function'' $f:\RR^n\to\RR^m$ becomes, in the quantum case, the map
\begin{equation}\label{eq:probability_vs_channel_and_rho}
    \bs p_{\Lambda,\mu}:
    \rho\mapsto
    \left(\Tr[\mu_b \Lambda(\rho)]\right)_{b=1}^{|\Sigma|}
    \in \RR^{|\Sigma|},
    % \langle \mu_b,\rho\rangle
\end{equation}
which sends each input state to the vector of outcome probabilities corresponding to a channel $\Lambda$ and measurement $\mu$ (here $|\Sigma|$ is the dimension of the set of measurement outcomes).
Finally, the trained model for QELMs consists of a linear function $W$ applied to the vector of outcome probabilities.
This means that, during training, the algorithm optimises the parameters $W$ so as to minimise the distance between $W \bs p_{\Lambda,\mu}(\rho_k^{\rm tr})$ and $\bs y_k^{\rm tr}$, for all the states and target vectors in the training dataset.
In~\cref{table:ELM_vs_QELM} we provide a schematic breakdown of the differences between ELMs and QELMs.

\begin{table}[tbp]
    \centering
    \begin{tabular}{lll}
        \toprule
        & {ELM} & {QELM}
        \\ \midrule
        training data & $\{(\bs x_k, \bs y_k)\}_k$  & $\{(\rho_k, \bs y_k)\}_k$
        \\ model to train & $\bs x\mapsto W f(\bs x)$ & $\rho\mapsto W\bs p_{\Lambda,\mu}(\rho)$
        \\ parameters to train & $W$ & $W$
        \\ cost function & $\|\bs y_k- W f(\bs x_k)\|_2$ & $\|\bs y_k- W \bs p_{\Lambda,\mu}(\rho_k)\|_2$
        \\ \bottomrule
    \end{tabular}
    \caption{Summary of the differences between classical ELMs and QELMs. These two schemes differ in the type of input states $\rho_k$ fed to the reservoir, and in how the reservoir map itself is implemented: in the classical case, is some nonlinear function often implemented via fixed-weights neural network architectures, whereas in the quantum case it is a quantum channel followed by some measurement.}
    \label{table:ELM_vs_QELM}
\end{table}

\parTitle{Formal description}
In the most general case, the channel $\Lambda$ is physically implemented by making $\rho$ interact with some \textit{reservoir state} $\eta$ and then tracing out some degrees of freedom from the output space.
This scenario can be modeled as a CPTP channel $\Phi\in\rmC(\calH_S\otimes\calH_E,\calH_E)$ sending states in $\calH_S\otimes\calH_E$ into states in $\calH_E$,
where $\calH_S$ and $\calH_E$ are the Hilbert spaces of input and reservoir states, respectively, and $\rmC(\calX,\calY)$ denotes the set of quantum channels sending states in $\calX$ to states in $\calY$.
For notational clarity, we will distinguish between the two channels $\Lambda_\eta\in\rmC(\calH_S,\calH_E)$ and $\calE_\rho\in\rmC(\calH_E,\calH_E)$, defined from $\Phi$ as
    $\Lambda_\eta(\rho) = \calE_\rho(\eta) = \Phi(\rho\otimes\eta),$
where $\eta$ and $\rho$ are states in $\calH_E$ an $\calH_S$, respectively.
Note that describing the channel as $\Phi\in\rmC(\calH_S\otimes\calH_E,\calH_E)$, means, in particular, that we assume the output space to have the same dimension as the input reservoir space. One could easily lift this restriction by considering measurements performed on the full space $\calH_S\otimes\calH_E$, nonetheless we stick to it as it eases our notation.

In the context of open quantum systems, dynamics through a reservoir are often described through channels acting on the reservoir itself, parametrised by the input state.
% \highlight{(qualche referenza?)}.
When adopting this point of view, the channel $\calE_\rho$ is the one of more direct interest.
% In such cases, one would consider the channel $\eta\mapsto \Phi(\rho\otimes\eta)$ for a given value of $\rho$.
This is useful for example when studying the memory capabilities of $\Phi$.
On the other hand, when one is interested in the retrievability of information encoded in $\rho$, the linearity of $\Lambda_\eta$ is of more direct relevance.

\subsection{Main results}

\parTitle{Linearity}
An observation central to our results is that the mapping from states to probabilities is, regardless of any detail of the dynamics, unavoidably linear
\begin{equation}
	\bs p_{\Lambda,\mu}(\alpha X+\beta Y)
	= \alpha \bs p_{\Lambda,\mu}(X)
	+ \beta \bs p_{\Lambda,\mu}(Y),
\end{equation}
for any pair of linear maps $X,Y$ and scalars $\alpha,\beta\in\CC$.
Furthermore, $\bs p_{\Lambda,\mu}(\rho)$ can be interpreted as a direct measure on the state $\rho$ --- that is, the overall process of measuring after an evolution $\Lambda$ can be reframed as an effective measurement performed directly on $\rho$. Explicitly, this follows from
\begin{equation}\label{eq:probability_with_effective_measurement}
	(\bs p_{\Lambda,\mu}(\rho))_b
	= \Tr[\mu_b\Lambda(\rho)]
	= \Trace[\Lambda^\dagger(\mu_b)\rho]
    = \Tr[\tilde\mu_b\rho],
\end{equation}
where $\Lambda^\dagger$ is the \textit{adjoint} of $\Lambda$, and $\tilde\mu_b$ denotes said effective measurement which, performed on $\rho$, reproduces the same measurement outcomes obtained measuring $\mu_b$ on $\Lambda(\rho)$.
One can equivalently view $\Lambda^\dagger(\mu_b)$ as describing the underlying evolution in the Heisenberg picture.
Because the measurement probabilities ultimately depend on the effective POVM $\tilde\mu$, we will use the shorthand notation $\bs p_{\tilde\mu}\equiv \bs p_{\Lambda,\mu}$ when $\tilde\mu=\Lambda^\dagger (\mu)$.

\parTitle{Characterisation of QELM-retrievable information}
A defining feature of QELMs is the restriction to \textit{linear} post-processing of the measurement probabilities, which has significant implications for their information processing capabilities.
To see this, note that applying the linear function $W$ to $\bs p_{\tilde\mu}(\rho)$ produces a vector $\bs y\equiv W\bs p_{\tilde\mu}(\rho)$, with components
\begin{equation}\small
    y_k {=} \sum_{b\in\Sigma} (\bs p_{\tilde\mu}(\rho))_b W_{kb}
    {=} \Tr[
        \left( \sum_{b\in\Sigma} W_{kb}\tilde\mu_b \right)
        \rho
    ].
    % = \Tr(\tilde\calO_k \rho).
\end{equation}
In other words, any vector $\bs y$ obtainable via linear post-processing of measurement probabilities has the form
% \begin{equation}
$y_k = \Tr(\tilde\calO_k \rho)\equiv\langle\tilde\calO_k,\rho\rangle$
% \end{equation}
for some observable $\tilde\calO_k$ that is a linear combination of the effective POVM elements. Here and in the following we use the notation $\langle\cdot,\cdot\rangle$ to highlight that expressions of the form $\Tr(A^\dagger B)$ can be interpreted as an inner product between the matrices.
It follows that a QELM can learn to retrieve the expectation value of an observable $\calO$ \textit{if and only if}
\begin{equation}\label{eq:condition_for_reconstructable_observables}
    \calO \in\on{span}_\RR (\{\tilde\mu_b: b\in\Sigma\}),
\end{equation}
that is, if and only if $\cal O$ can be written as a real linear combination of operators $\tilde\mu_{b}$.
% More explicitly, this condition means that $\calO$ can be written as a real linear combination of the operators $\tilde\mu_b$.
It is worth noting that, in this context, we operate under the assumption that $\tilde\mu$ --- and thus $\Lambda$ and $\mu$ --- is known, and therefore the condition is readily verifiable.
In particular, a QELM can reproduce the expectation value of arbitrary observables \textit{iff} $\tilde\mu$ is \textit{informationally complete} (that is, \textit{iff} $\{\tilde\mu_b:b\in\Sigma\}$ spans the corresponding space of Hermitian operators).
Nonetheless, as will be further discussed later, the training procedure does not require knowledge of $\tilde\mu$ as it can be seen as a way to \textit{estimate} the effective measurement $\tilde\mu$ itself.

\subsection{Reconstruction method}

\parTitle{How to retrieve the information in practice}
Even if we can now readily assess whether a target observable can be retrieved from the information provided in a given QELM setup, the question remains on how exactly this would be done.
To fix the ideas, consider a scenario with a single target observable $\calO$, and the effective POVM is some $\bstildemu$ with $|\bstildemu|$ the number of possible outcomes.
The problem is thus finding some $W$ --- which will be, in this case, a row vector --- such that
\begin{equation}\label{eq:main_problem_statement}
    \langle\calO,\rho\rangle = W\langle\bstildemu,\rho\rangle
\end{equation}
for all the elements of the training dataset, which has in this case the form
% \highlight{(Ivan: se si cambia la $\bs p$ in $\rho$ di pag 2, va cambiata anche la 7)}
\begin{equation}
    \{(\langle\calO,\rho\rangle,\langle\bstildemu,\rho\rangle): \rho\in\mathrm{TrainingDS}\},
\end{equation}
where $\mathrm{TrainingDS}$ is the set of states used to generate the training dataset.
A convenient way to write this condition is then
\begin{equation}\label{eq:main_training_equation}
    \langle\calO,\rho^{\rm tr}\rangle =
    W \langle\bstildemu,\rho^{\rm tr}\rangle,
\end{equation}
denoting with $\rho^{\rm tr}$ the vector whose elements are all the training states, with $\langle\calO,\rho^{\rm tr}\rangle$ the vector of expectation values $\langle\calO,\rho_k^{\rm tr}\rangle$, and with $\langle\bstildemu,\rho^{\rm tr}\rangle$ the matrix with components $\langle\tilde\mu_b,\rho_k^{\rm tr}\rangle$.
\Cref{eq:main_training_equation}, as a condition for $W$, is a standard linear regression problem.
It is however worth remarking a departure of our task from standard linear regressions: we are not interested in finding any ``true value'' of $W$, but rather in finding {\it some} $W$ which gives the best performances on the test dataset.
That means, in particular, that the existence of multiple optimal solutions for $W$ is not an issue.

\parTitle{Training as POVM reconstruction}
In the context of QELM, the effective measurement $\bstildemu$ --- and thus the matrix $\langle\tilde\mu,\sigma\rangle$ --- is not known \textit{a priori}.
Instead, during the training phase, only the probabilities  $\langle\bstildemu,\bs\rho^{\rm tr}\rangle$ and expectation values $\langle\bs\calO,\bs\rho^{\rm tr}\rangle$ are given.
The task is to solve the corresponding linear system
\begin{equation}\label{eq:linear_system_defining_W}
    \langle\bs\calO,\bs\rho^{\rm tr}\rangle =
    W \langle \bstildemu,\bs\rho^{\rm tr}\rangle
\end{equation}
for $W$.
Even though without knowing $\bs\calO$ and $\bstildemu$ it is not possible to determine \textit{a priori} the feasibility of the task, if the accuracies during the training phase are sufficiently high one can reasonably expect the condition to be fullfilled.
If, on the other hand, the accuracies saturate to a non-optimal amount while increasing the sampling statistics, we can now determine the reason to be $\bs\calO$ not being writable as linear combinations of $\bstildemu$.

A standard way to solve~\cref{eq:linear_system_defining_W} is via the pseudoinverse
% \begin{equation}
%     W =
%     \langle\bs\calO,\bs\rho^{\rm tr}\rangle
%     \langle\bstildemu,\bs\rho^{\rm tr}\rangle^+,
% \end{equation}
\begin{equation}\label{eq:pseudoinverse_solution_for_W}
\begin{aligned}
    W &=
    \langle\bs\calO,\bs\rho^{\rm tr}\rangle
    \langle\bstildemu,\bs\rho^{\rm tr}\rangle^+ %\\
%&   = \langle\bs\calO,\bs\rho^{\rm tr}\rangle
  %  (
   %  \langle\bs\rho^{\rm tr},\bstildemu\rangle
   %  \langle\bstildemu,\bs\rho^{\rm tr}\rangle
   % )^{-1}
   % \langle \bs\rho^{\rm tr},\bstildemu\rangle.
\end{aligned}
\end{equation}
where $A^+$ denotes the pseudoinverse of $A$.
This solution is exact \textit{iff} $\on{supp}(\langle\calO,\rho^{\rm tr}\rangle)\subseteq \on{supp}(\langle\tilde\mu,\rho^{\rm tr}\rangle)$, and unique \textit{iff}
$\on{supp}(\langle\calO,\rho^{\rm tr}\rangle)= \on{supp}(\langle\tilde\mu,\rho^{\rm tr}\rangle)$~\cite{higham2002accuracy}.
Given a Hermitian operator $X$ and an informationally complete POVM $\bstildemu$, there is always a \textit{dual} POVM $\bstildemu^\star$, with $|\bstildemu|=|\bstildemu^\star|$ that allows the decomposition~\cite{casazza2015brief}
\begin{equation}
   X
    = \sum_k \langle \tilde\mu_k^\star,X\rangle \tilde\mu_k
    = \sum_k \langle \tilde\mu_k,X\rangle \tilde\mu_k^\star.
\end{equation}
The POVM $\bstildemu^\star$ is also referred to, in this context, as a \textit{dual frame} of $\bstildemu$.
A particular choice of such a dual basis is constructed as
\begin{equation}
    \tilde\mu^\star_k = S^{-1}(\tilde \mu_k),
    \quad
    S(X) \equiv \sum_k \tilde\mu_k \langle\tilde\mu_k,X\rangle,
\end{equation}
where $S$ is referred to as the \textit{frame operator}, which is ensured to be invertible, provided $\bstildemu$ is informationally complete, and this basis is the \textit{canonical dual frame} of $\bstildemu$.
With $\bstildemu^\star$, we can write
\begin{equation}
    \langle\calO_i,\rho\rangle
    = \sum_k \langle \calO_i,\tilde\mu^\star_k\rangle
    \langle\tilde\mu_k,\rho\rangle,
\end{equation}
which tells us that a general solution to the linear reconstruction problem has the form
\begin{equation}
    W = \langle\bs\calO,\bstildemu^\star\rangle.
\end{equation}
This provides a very concrete understanding of what the training phase achieves: through training, and solving the associated linear problem, we retrieve a partial description of the measurement process itself, through its dual operators. Note that one can also consider this framework using a complete set of observables $\calO_i$ as target, in which case $\langle\bs\calO,\bstildemu^\star\rangle$ also amounts to a complete characterisation of $\bstildemu^\star$, and thus of $\bstildemu$.

\parTitle{Test phase and accuracy}
The performance of the QELM is quantified by its accuracy on previously unseen ``test'' states.
A standard choice of quantifier is the \textit{mean squared error} (MSE): given a test state $\rho$, and assuming that the training produced parameters $\bs w$, this reads
\begin{equation}
\text{MSE}=    (\langle\calO,\rho\rangle -
    \bs w\cdot \langle\bstildemu,\rho\rangle)^2.
\end{equation}
For multiple target observables, the definition is extended straightforwardly: we have
\begin{equation}
\text{MSE}=    \|
    \langle\bs\calO,\rho\rangle -
    W \langle\bstildemu,\rho\rangle
    \|_2^2,
\end{equation}
where now $\bs\calO=(\calO_1,\calO_2,...)$ is a vector of target observables, and $W$ the matrix obtained from the training phase.

\parTitle{Effects of finite statistics}
In an ideal scenario, where the probabilities $\langle\bstildemu,\rho\rangle$ are known with perfect accuracy, solving~\cref{eq:main_training_equation} is not an issue. Assuming that the system is indeed solvable --- that is,~\cref{eq:condition_for_reconstructable_observables} is satisfied --- then any solution method, \textit{e.g.} computing the pseudo-inverse of $\langle\bstildemu,\rho^{\rm tr}\rangle$, will result in some $W$ which maps perfectly well measurement probabilities to expectation values.
However, any realistic scenario will result in a radically different outlook.
Because the protocol uses measurement probabilities as fundamental building blocks, being mindful of potential numerical instabilities is paramount.
In particular, the probability vectors $\langle\bstildemu,\rho\rangle$ will only be known up to a finite accuracy which depends on the finite number $N$ of statistical samples, since the variance of the estimates will scale as $N^{-1}$.

These statistical fluctuations will both affect the estimation of $\langle\bs\calO,\bstildemu^\star\rangle$ in the training phase, and the final accuracies in the testing phase.
% , due to the combined effect of the errors in the estimated $\langle\bs\calO,\bstildemu^\star\rangle$, and the noise in the measurement probabilities used for the estimation itself.
The latter source of noise is present even if $\langle\bs\calO,\bstildemu^\star\rangle$ is known with perfect accuracy, while the former is due to the use of a finite training dataset.

\subsection{Reconstruction efficiency }

\parTitle{Ill-conditioning}
An important factor to consider when using QELMs is the potential numerical instability arising from solving the associated linear system~\cite{higham2002accuracy}.
While~\cref{eq:pseudoinverse_solution_for_W} provides a general and efficiently computable solution to the learning problem, this solution can be \textit{ill-conditioned}, i.e. %Roughly speaking, a problem is said to be ill-conditioned when 
small perturbations of the inputs can result in large perturbations of the outputs.
In our context, this happens when $\langle\tilde\mu,\rho^{\rm tr}\rangle$ has small singular values, which might arise due to noise or finite statistics.
The issues associated to solving a linear system in a supervised learning context, and some possible ways to tackle them, are discussed in~\cite{rosasco2004learning,devito2005learning,zhao2009improving}.
Depending on the circumstances, several regularisation techniques can be used to deal with ill-conditioned problems.

A standard way to quantify the potential ill-conditioned nature of a linear system is the \textit{condition number}~\cite{higham2002accuracy}: Given a linear problem $\bs y=A\bs x$ which one wishes to solve for $\bs x$, the \textit{condition number} of $A$ is 
$$\kappa(A)=\frac{s_{\rm max}}{s_{\rm min}},$$ 
where $s_{\rm max}$ ($s_{\rm min}$) is the largest (smallest) singular values of $A$.
The set of solutions to the linear system is the affine space
\begin{equation}
    \bs x \in A^+ \bs y + \ker(A),
\end{equation}
where $A^+$ denotes the \textit{pseudo-inverse} of $A$.
A simple characterisation of $\kappa(A)$ is that it provides the worst-case scenario estimate of relative error amplification: if $\Delta\bs y$ is the error associated with $\bs y$, the relative error on $\bs x$ is bounded by
\begin{equation}\label{eq:error propagation}
    \left\|\frac{\Delta\bs x}{\bs x}\right\| \le
    \kappa(A) \left\|\frac{\Delta\bs y}{\bs y}\right\|.
\end{equation}
\Cref{eq:linear_system_defining_W} is precisely the type of linear system whose numerical stability is estimated via the condition number, in this case $\kappa(\langle\bstildemu,\bs\rho^{\rm tr}\rangle)$, the overarching goal of QELMs is not accurately estimating $W$, but rather finding any $W$ that results in accurately estimating the target expectation values on the test dataset.
In other words, we only care about inaccuracies in the estimation of $W$ in so far as they are reflected in inaccuracies in the MSE
$\|\langle\bs\calO,\rho\rangle - W\langle\bstildemu,\rho\rangle\|_2$.
That means the errors we are interested in are those coming from the expression
\begin{equation}
    \underbrace{(\langle\bs\calO,\bs\rho^{\rm tr}\rangle
    \langle\bstildemu,\bs\rho^{\rm tr}\rangle^+)}_{=W}
    \langle\bstildemu, \rho\rangle,
\end{equation}
where both $\langle\bstildemu,\bs\rho^{\rm tr}\rangle$ and $\langle\bstildemu, \rho\rangle$ are estimated up to some finite precision.

\parTitle{Ill-conditioning due to statistical noise}
An unavoidable source of ill-conditioning is the fundamental statistical nature of the probabilities entering the $|\bstildemu|\times M_{\rm tr}$ matrix $P\equiv \langle\bstildemu,\bs\rho^{\rm tr}\rangle$.
Let $P_N$ denote the matrix whose elements are the frequencies associated with the corresponding probabilities in $P$, estimated from $N$ samples.
% This can thus be thought of as a matrix-valued random variable.
If the input states have dimension $N_{\rm input}$ (\textit{e.g.} $N_{\rm input}=4$ for 2 qubits), but $|\bstildemu|>N_{\rm input}$, then $P$ will have some vanishing singular values.
Due to the statistical noise, these will become nonzero, albeit remaining relatively small with magnitude of the order of $1/N$, in $P_N$.
This makes the linear inversion problem potentially ill-conditioned, as the eigenspaces corresponding to such singular values do not represent physically relevant information.
A simple way to fix this issue is to truncate the singular values, setting to zero those beyond the $N_{\rm input}$-th one.
This strategy does not introduce a significant amount of error, as long as the variances associated to the outcome probabilities are sufficiently smaller than all the other (physically relevant) singular values, which is always the case for sufficiently large $N$.
We will employ this strategy for our simulations.

\begin{figure*}[t]
	\includegraphics[width=\linewidth]{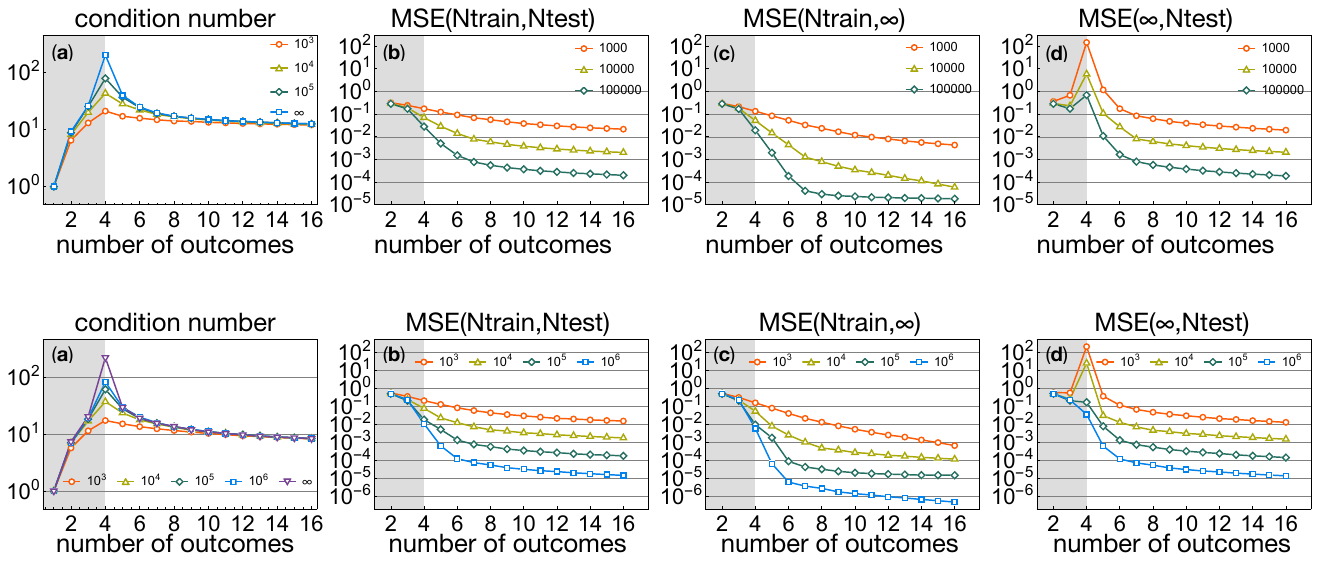}
	\caption{
	\textit{Mean squared error (MSE) associated to reconstruction of $\Tr(\calO\rho)$ for some single-qubit observable $\calO$ in the first scenario configuration, with $M_{\rm tr}=100$ and $M_{\rm test}=1000$ states used during training and testing phase, respectively.}
	In all plots, different colours refer to different numbers of samples $N_{\text{train}},N_{\text{test}}$ used to estimate the probabilities.
	The target observable is chosen at random, and kept fixed in all shown simulations. Choosing different observables does not significantly affect the behaviour of these plots. 
	\textbf{(a)} Condition number of the probability matrix $\langle\bstildemu,\bs\rho^{\rm tr}\rangle$ as a function of the number of measurement outcomes.
	\textbf{(b)} MSE as a function of the number of measurement outcomes, when both train and test probabilities are estimated with the same finite precision.
	\textbf{(c)} As above, but now the test probabilities are estimated with infinite precision.
	\textbf{(d)} As above, but now the training probabilities are estimated with infinite precision.
    In this last case, the large error corresponding to four outcomes is due to the amplification of the statistical error in the vector of probabilities $\bs{p}$ by the map $W$. The amount of  amplification is described by the condition number in~\cref{eq:error propagation}.
	} 
	\label{fig:MSE_vs_numMeas_linear}
\end{figure*}

\parTitle{Dependence of condition number on statistics}
Another interesting feature is the increase of the condition number $\kappa(P_N)$ on $N$ [cf.~\cref{fig:MSE_vs_numMeas_linear}-\textbf{(a)}]. This is somewhat counterintuitive, as we would expect estimation to become easier when the probabilities are known more accurately. 
We refer to~\cref{A1} for a detailed discussion of this aspect.

\section{Single-injection examples}
\label{sec:single-inj-examples}

\parTitle{Types of dynamics considered}
Let us consider how our framework applies to the case with single-qubit inputs.
Most of the literature focuses on reservoir dynamics defined via some Hamiltonian~\cite{fujii2017harnessing}, or on open quantum systems~\cite{ghosh2019quantum}. Our aim is here to study the performance of QELMs in standard scenarios, and we therefore focus on unitary evolutions, and analyze cases where the reservoir dynamics is a random unitary or isometric evolution rather than a specific Hamiltonian model, in order to gain a better insight into the performances of QELMs in more general contexts.
More specifically, we focus on the following three scenarios:
\begin{enumerate}
    \item The input qubits interact with a high-dimensional state through some random unitary evolution. In this case, the reservoir is a qudit, measured in some fixed computational basis, and the corresponding evolution reads: $\Lambda(\rho) = \Tr_1[V\rho V^\dagger]$ with $V\in\rmU(\CC^2,\CC^2\otimes \CC^n)$ a $(2n)\times2$ isometry, for some $\NN\ni n>2$.
    In this notation, the initial state of the reservoir is implicitly specified through the choice of isometry $V$.
    The corresponding measurement is taken to be $\mu_j=\ketbra j$ with $j=1,..., n$, and the effective measurement thus reads
    \begin{equation}\label{eq:POVM_unitary_evolution}
        \tilde\mu_j = \Lambda^\dagger(\ketbra j)
        = V^\dagger(I\otimes \ketbra j)V.
    \end{equation}
    \item Alternatively, one can consider a scenario involving a single high-dimensional qudit, with no bipartite structure involved.
    In this case, the ``input qubit'' is a two-dimensional subspace of the qudit, $\tilde\rho=\rho\oplus \eta_0$ with $\rho\in\rmD(\CC^2)$ a single-qubit state, and $\eta_0\in\rmD(\CC^{n-2})$ the initial state of the reservoir degrees of freedom.
    The dynamics is in this case simply an evolution of the form $\rho\oplus\eta_0\mapsto U (\rho\oplus\eta_0) U^\dagger$ for some unitary operator $U\in\rmU(\CC^n)$.
    Measurements are again performed in the computational basis, $\mu_k=\ketbra k$ with $k=1,...,n$, and thus
    \begin{equation}
        \tilde\mu_k = U^\dagger\ketbra k U.
    \end{equation}
    In this notation, the degrees of freedom of the input state are also measured after the evolution, but this is not an issue for our purposes.
    \item As a further example, let us consider a system of qubits interacting through some Hamiltonian $H$. In this case, the input qubit interacts with $N_R$ reservoir qubits through some Hamiltonian, and the measurement is performed on the reservoir qubits.
    The dynamics thus has the form
    \begin{equation}
        \rho \mapsto e^{-iHt} (\rho\otimes\eta_0) e^{iHt}
    \end{equation}
    for some evolution time $t$ and initial reservoir state $\eta_0\in\rmD(\CC^{2^{N_R}})$. For our tests, we use a pairwise Hamiltonian for a qubit network of the form
    \begin{equation}
        H = \sum_{ij=1}^{N_R+1} J_{ij}(\sigma^+_i\sigma^-_j+\sigma^+_j\sigma^-_i)+\sum_{i=1}^{N_R}\Delta_i \sigma_i^x,
        \label{eq:hamiltonian_example}
    \end{equation}
    % \highlight{(and the other types of dynamics we consider, chain etc)}
    with random coupling constants $J_{ij}$ drawn uniformly at random from the interval $[-1,1]$, and driving coefficients $\Delta_i$ drawn uniformly at random from $[0,1]$.
    We consider different network connectivities; in particular we study (1) a linear chain with nearest-neighbor interactions, (2) a fully connected reservoir, with a single node connected to the input, and finally (3) a fully connected reservoir where each node is connected to the input.
    If measurements are again performed in the computational basis of the reservoir, that is $\mu_k=\ketbra k$ with $k=1,...,2^{N_R}$, the corresponding effective measurements will have the form
    \begin{equation}
    \label{eq:effective_POVM_hamiltonian_dyn}
        \tilde\mu_k = \Tr_2\left[(I\otimes \eta_0) e^{iHt} (I\otimes \ketbra k) e^{-iHt}\right].
    \end{equation}
\end{enumerate}

As training objective, we consider the reconstruction of the expectation value of some target observable $\calO\in\Herm(\CC^2)$.
For our simulations, we make the conventional choice $\calO=\sigma_x$, with $\sigma_x$ the Pauli $X$ matrix.
Note that choosing different observables or different evolutions,
does not significantly affect the results.

To reconstruct arbitrary linear functionals of $\rho$, the effective measurement must have rank four, that is, it must contain four linearly independent operators. This is required to have tomographically complete knowledge of $\rho$.
This means that, in particular, the reservoir state must be at least four-dimensional.

\begin{figure*}[t]
	\includegraphics[width=\linewidth]{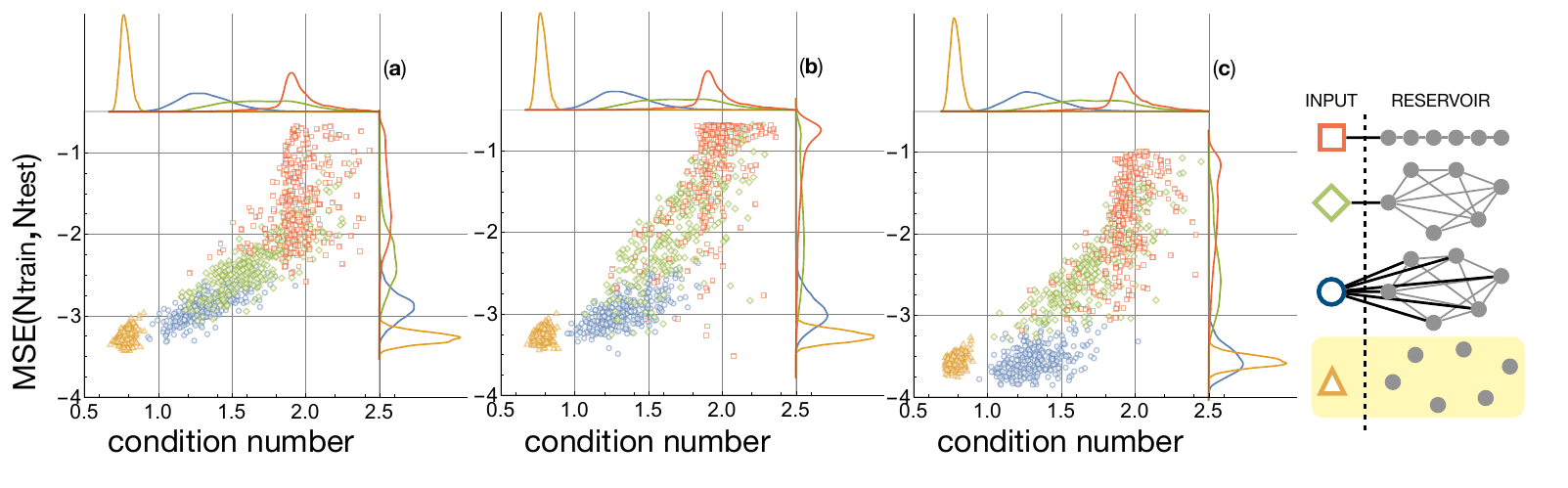}
	\caption{
	\textit{MSE (in logarithmic scale) obtained by training random reservoirs corresponding to different types of dynamics to retrieve a fixed target observable, shown against the condition number.}
	The target one-qubit observable $\calO$ is \textbf{(a)} the $\sigma_x$ Pauli matrix, \textbf{(b)} the $\sigma_z$ Pauli matrix, and \textbf{(c)} a one-qubit observable sampled at random.
	In each case, we plot data corresponding to a reservoir dynamics that is (red squares) a random one-dimensional spin chain with nearest neighbor interactions, (green diamond) a random fully connected spin Hamiltonian, where the input is only connected to a single node of the reservoir, (blue circles) a random fully connected spin Hamiltonian, and (orange triangles) a random unitary evolution.
	Each point shows simulation results obtained using $M_{\rm tr} = 100$ training states, $M_{\rm test}=1000$ test states, and a reservoir comprised of 6 qubits.
    The statistics is fixed to $N_{\text{train}}=N_{\text{test}}=10^4$ samples used to estimate each measurement probability. Except for the random unitary case (orange triangles), representative of the first scenario, the other configuration are examples of the third scenario and the Hamiltonian parameters $J_{ij}$ and $\Delta_i$ from \cref{eq:hamiltonian_example} are sampled uniformly at random in the interval $[0,1]$.
	}
	\label{fig:nuvole_MSE_CN}
\end{figure*}

\Cref{fig:MSE_vs_numMeas_linear} reports the performances of QELMs trained to retrieve $\sigma_x$, when the evolution corresponds to an input qubit interacting with a $2^5$-dimensional qudit through a random unitary operator, for different numbers of elements in the effective POVM $\{\tilde\mu_k\}$. 
Let $|\tilde\mu|$ denote the number of such elements.
In the ideal scenario where training and test probabilities are known with perfect accuracy, the MSE is precisely zero whenever $|\tilde\mu|\ge4$.
To get more realistic results, we consider the performance when $\langle\tilde\mu,\rho^{\rm tr}\rangle$ and $\langle\tilde\mu,\rho^{\rm test}\rangle$ are estimated from finite statistics.
In these scenarios, the condition number of the matrix $\langle\tilde\mu,\rho^{\rm tr}\rangle$ is also relevant, as it correlates with how much the statistical fluctuations in $\langle\tilde\mu,\rho^{\rm test}\rangle$ can be amplified and lead to estimation inaccuracies.
As shown in the figures, the accuracy increases with better statistics, as expected, but also when increasing $|\bstildemu|$.
It is worth stressing that this feature does not occur with the ideal probabilities, as in that scenario the MSE is perfectly zero from four measurements onwards~\footnote{More precisely, we should say that the ideal MSE vanishes \textit{almost always} when the unitary evolution is drawn uniformly at random. It is in fact possible to find examples of unitaries which make the reconstruction impossible. Trivial examples would be unitaries that do not correlate input and reservoir degrees of freedom. These cases almost never occur when drawing unitaries at random, however}.

\Cref{fig:MSE_vs_numMeas_linear} shows that, although 4 measurements are in principle sufficient to retrieve the target information, reconstruction in realistic circumstances becomes easier when increasing the dimension of the reservoir, that is, the number of measurement outcomes.
In \cref{fig:MSE_vs_numMeas_linear}-\textbf{a} we see that the numerical problem becomes better conditioned when there are more measurement outcomes. %, as quantified by the decreasing condition number. 
In~\cref{fig:MSE_vs_numMeas_linear}-\textbf{b} and \cref{fig:MSE_vs_numMeas_linear}-\textbf{c} we appreciate how the accuracy increases when more statistical samples are used, and thus the probabilities approach  their ideal values.

\Cref{fig:MSE_vs_numMeas_linear}-\textbf{d} shows the MSE when the training parameters are computed from the ideal probability matrix $\langle\tilde\mu,\rho^{\rm tr}\rangle$, while finite statistics is used in the testing phase. In this case,  the poor statistical accuracy found for small numbers of outcomes is due to the correspondingly large condition number.
This is to be attributed to numerical instability associated with the ideal reconstruction parameters for few measurement outcomes: indeed the large error corresponding to four outcomes is due to the amplification of the statistical error in the vector of probabilities, amplification that is quantified by the condition number  \cref{eq:error propagation}.
Note that such detrimental effect largely disappears already for $|\tilde\mu|\ge8$.
Note that the data shown in~\cref{fig:MSE_vs_numMeas_linear}-\textbf{(d)} and in the purple triangles in \cref{fig:MSE_vs_numMeas_linear}-\textbf{(a)} corresponds to a training performed with perfectly estimated training probabilities. Even if not directly related to performances in practical scenarios, this data is useful to better isolate the different effects caused by inaccuracies during training and testing phases.

Finally, in~\cref{fig:nuvole_MSE_CN} we consider how different choices of dynamics influence the reconstruction performances. In particular, we consider input states interacting with the reservoir through a random unitary evolution, a random pairwise Hamiltonian, or a randomly drawn pairwise Hamiltonian with a chain structure, in which each qubit only interacts with its nearest neighbour. 
Overall, as the degree of connectivity of the network increases, the performance of the reservoir and stability of the linear regression both improve. This is illustrated by the decrease in the MSE and the condition number.

\begin{figure}[t]
	\includegraphics[width=\linewidth]{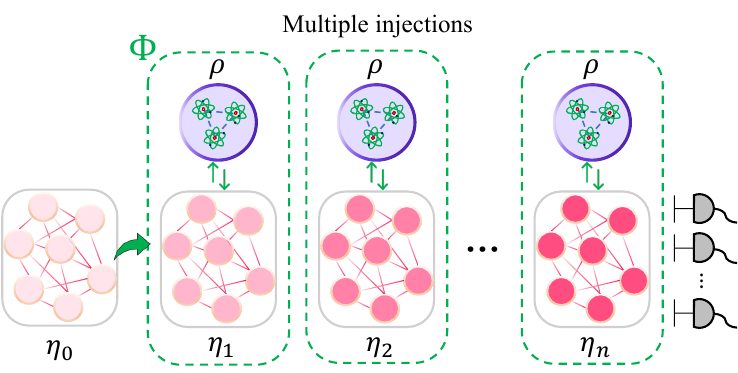}
	\caption{
	    Schematics of the QELM protocol in the multiple injections configuration: the reservoir interacts with multiple copies of the same quantum state $\rho$ and progressively acquires information about it.
	    The measurement is performed on the final reservoir state $\eta_n$.
	}
	\label{fig:scheme_QELM_vs_QRCSs}
\end{figure}

\section{Multiple injections}
\label{sec:QELM_multiple_injections}

In~\Cref{sec:QELM_single_injection} we focused on the achievability of target observables when single copies of an input state are made to interact with a reservoir, which is then measured.
As shown, this characterises the amount of exactly retrievable information from functionals that are linear in the input density matrix.
In this Section we consider the more general scenario where \textit{several copies} of an input state are used as input.
This allows us to retrieve a broader range of properties of the input states.

\subsection{Main results}
\parTitle{Possible targets with multiple injections}
Consider a channel $\calE_\rho$ applied \textit{multiple times} to an initial reservoir state $\eta_0$~\footnote{It is worth stressing here that there is no cloning involved in this process. The multiple injections are to be achieved by preparing the same state multiple times, not by cloning a single copy of the state.}. 
For $n$ consecutive uses of the channel and initial reservoir state $\eta_0$, the measured state is then $\eta_n=\calE_\rho^n(\eta_0)$.
This can be rewritten as
\begin{equation}
    \eta_n = 
    \Phi(\rho\otimes\Phi(\rho\otimes \eta_{n-2}))
    = \dots 
    = \tilde\Phi(\rho^{\otimes n}\otimes\eta_0),
\end{equation}
where we have introduced the resulting channel $\tilde\Phi$. By the argument used in~\Cref{sec:QELM_single_injection}, the possible outputs after linear post-processing of the outcome probabilities are all and only those of the form
\begin{equation}
    y = \Tr(\tilde{\calO} \rho^{\otimes n}),
\end{equation}
for some observable $\tilde{\calO}$ acting in the space of $n$ copies of $\rho$.

Training these models thus proceeds similarly to the linear case: the probabilities are estimated from measurements performed after each series of $n$ injections, and these probabilities are then used to solve~\cref{eq:main_problem_statement} and thus find the optimal train parameters $W$. ~\Cref{fig:scheme_QELM_vs_QRCSs} shows a scheme of the multiple-injection model here described. 

After $n$ injections, the space on which the effective POVM acts has dimension
\begin{equation}\label{eq:dimension_multiple_injection_space}
    d_{n,m}\equiv \binom{m^2+n-1}{n}
\end{equation}
with $\dim(\calH)\equiv m$ the dimension of each input state. This is the number of degrees of freedom characterising a symmetric tensor of the form $\rho^{\otimes n}$ with $\rho\in\rmD(\calH)$.
This is also the space where the target observables $\calO$ live. It follows that, in order to be able to reconstruct arbitrary functionals of $\rho$ up to the maximum order of $n$, the measurement must contain at least $d_{n,m}$ linearly independent components.

\parTitle{Purity example}
Consider for example the task of estimating the purity of a given state. Observe that the map $\rho\mapsto\Tr(\rho^2)$ can be written as $\Tr(\rho^2)= \Tr[\on{SWAP} (\rho\otimes\rho)]$.
As per our previous observations, this means that the purity can be retrieved from a QELM provided that at least two injections are used, and that the effective measurement $\tilde\mu$ is such that $\on{SWAP}$ can be expressed as a real linear combination of the measurement operators $\tilde\mu_b$. 

\subsection{Application examples}
\begin{figure}[t]
	\includegraphics[width=\linewidth]{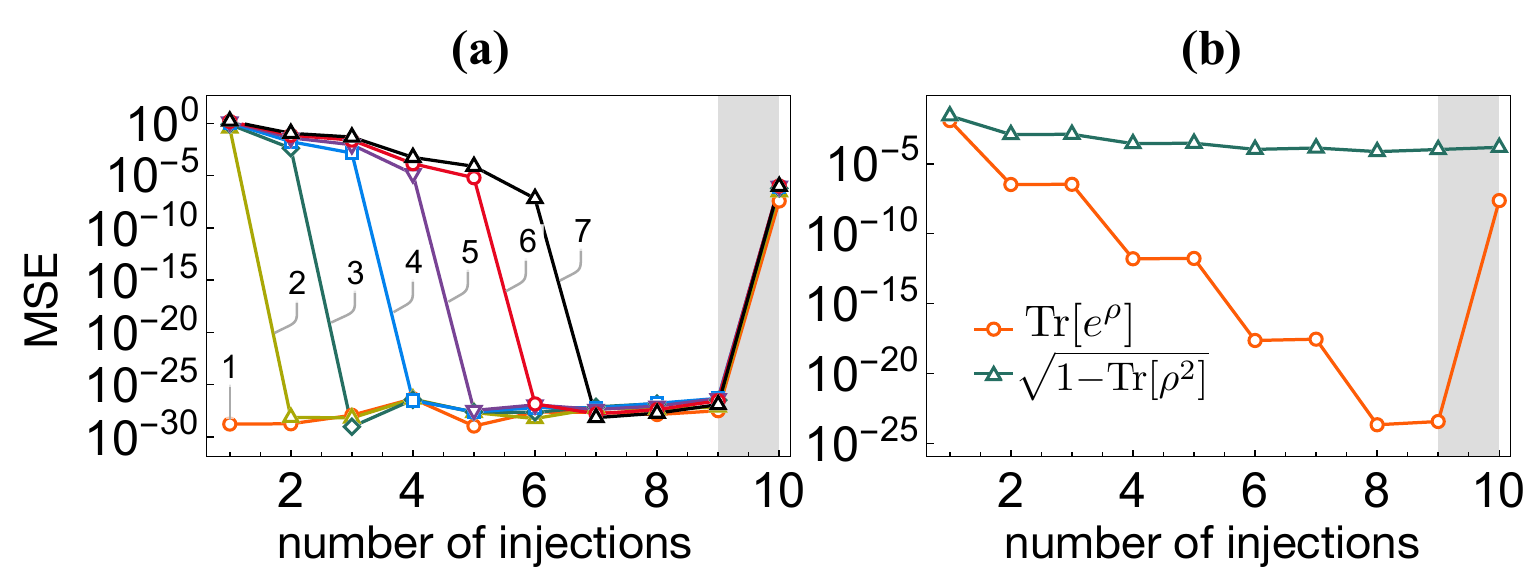}
	\caption{
	\textbf{(a)} Reconstruction MSE for polynomial targets $\Tr(\calO\rho^k)$ with $k=1,\dots,7$.
	As previously discussed, reconstruction is not possible unless the number of injections is greater than or equal to the degree of the target function (for polynomial target functions). At the same time, the number of independent measurements bounds the number of injections that can be used without degrading the information. Here, the reservoir consists of $8$ qubits, which is why the reconstruction fails when more than $9$ injections are used, consistently with~\cref{eq:dimension_multiple_injection_space}: with $8$ qubits there are $2^8=256$ available measurements, while the $9$-injection space has dimension $\binom{2^2+9-1}{9}=220<256$. Using $10$ injections implies a dimension $\binom{2^2+10-1}{10}=286>256$. 
	\textbf{(b)} Reconstruction MSE for the nonlinear targets $\Tr(e^\rho)$ and $\sqrt{1-\Tr(\rho^2)}$.
	Due to the nonlinearity of these functions, ideal reconstruction is never feasible. The significantly better performances obtained for $\Tr(e^\rho)$ are due to the coefficients of its Taylor series vanishing faster than those of $\sqrt{1-\Tr(\rho^2)}$.	}
	\label{fig:MSE_vs_numMeas_nonlinear}
\end{figure}
To showcase the reconstruction of nonlinear functionals of the input state,  in~\cref{fig:MSE_vs_numMeas_nonlinear} we consider targets functionals of the form $\Tr(\calO \rho^k)$, $\Tr(e^\rho)$, and $\sqrt{1-\Tr(\rho^2)}$.
% as well as the concurrence of input two-qubit states $C(\rho)$.
We focus on the number of injections required for the reconstruction in each case, and thus assume ideal training and target probabilities. 

In~\cref{fig:MSE_vs_numMeas_nonlinear}-\textbf{a} we give the MSE associated with the reconstruction of $\Tr(\calO \rho^k)$, $k=1,\dots,7$, for a random one-qubit observable $\calO$, for different numbers of injections. In these simulations, the reservoir is an $8$-qubit system, with an additional qubit used for the input states, reset to $\rho$ for each injection.
As expected from our previous discussion, we observe that the reconstruction is only successful when the number of injections $n$ is larger than the degree $k$ of the polynomial of the target observable.

Furthermore, note how the reconstruction fails again when the number of injections increases too much.
This \textit{upper bound} for the reconstruction is due to the finite dimension of the reservoir --- or, equivalently, the finite number of measurement operators in $\tilde\mu$.
In fact, reconstructing $\Tr(\calO\rho^k)$ from a measurement performed after $n$ injections amounts to reconstructing a specific observable $\tilde\calO$ acting on the space of states of the form $\rho^{\otimes n}$.
If the measurements are not suitably chosen, as is the case in QELM-like scenarios, this means that the number of (linearly independent) measurements must be sufficient to reconstruct all possible observables on such a space, whose dimensionality is $d_{n,m}$.

In~\cref{fig:MSE_vs_numMeas_nonlinear}-\textbf{(b)} we treat the case of non linear functionals of $\rho$. The performance achieved in approximating $\sqrt{1-\Tr(\rho^2)}$ is poor due to the slow convergence of the Taylor expansion of the functional. The step-like behavior that is evident in the MSE associated with the reconstruction of $\Tr(e^\rho)$, which is also present in the case of $\sqrt{1-\Tr(\rho^2)}$ although less evidently, can be explained by noticing that the trace of odd powers of $\rho$ is a polynomial of the same degree of the previous even ones.

\section{Conclusions} \label{conc}

We provided a complete characterisation of the information exactly retrievable from linear post-processing of measurement probabilities in QELM schemes.
This  sheds light on the tight relation between the capabilities of a device to retrieve nonlinear functionals of input states, and the memory of the associated quantum channel.

We found that the estimation efficiency of QELM protocols is entirely reflected in the properties of an effective POVM describing the entire apparatus, comprised of a dynamical evolution and a measurement stage.
\addLI{In particular, we showed that the effective POVM contains all of the information required to determine which observables can be estimated, and to what accuracy, as well as which kinds of effective POVMs, induced by different types of dynamics, result in different degrees of estimation accuracies.}
In turn, this clarifies the class of dynamics that result in POVMs that are effective for efficient and accurate property estimations. We further found that the inevitable sampling noise, intrinsic to any measurement data coming from a quantum device, crucially affects estimation performances, and cannot be neglected when discussing the protocols.

Our work paves the way for a number of interesting future endeavours on this line of research, including an extension of our analysis to time-trace signals for dynamical  QRCs, and the in-depth analysis of POVM optimality for quantum state estimation. Moreover, the translation of our findings into performance-limiting factors of recently designed experimental scenarios for QELMs and QRCs, and the identifications of ways to counter them, will be paramount for the grounding of the role that such architectures could play in the development of schemes for quantum property validation.
\addLI{At the same time, our study of QELMs for quantum state estimation purposes fits tightly with, and has the potential to improve on, several experimental detection strategies which rely on some form of linear regression to estimate target states~\cite{zia2023regression,suprano2021enhanced,suprano2021dynamical,stricker2022experimental,garcia2021learning}.}
\\

\acknowledgments

LI acknowledges support from MUR and AWS under project 
PON Ricerca e Innovazione 2014-2020, ``calcolo quantistico in dispositivi quantistici rumorosi nel regime di scala intermedia" (NISQ - Noisy, Intermediate-Scale Quantum). IP is grateful to the MSCA Cofund project CITI-GENS (Grant nr. 945231). MP acknowledges the support by the European Union's Horizon 2020 FET-Open project  TEQ (766900),  the Horizon Europe EIC Pathfinder project QuCoM (Grant Agreement No.\,101046973), the Leverhulme Trust Research Project Grant UltraQuTe (grant RGP-2018-266), the Royal Society Wolfson Fellowship (RSWF/R3/183013), the UK EPSRC (EP/T028424/1), and the Department for the Economy Northern Ireland under the US-Ireland R\&D Partnership Programme.

\bibliography{main}
% \printbibliography

% \newpage
\appendix
\section{Under estimation of the condition number}
\label{A1}
\begin{figure}[htb]
    \centering
    \includegraphics[width=0.9\linewidth]{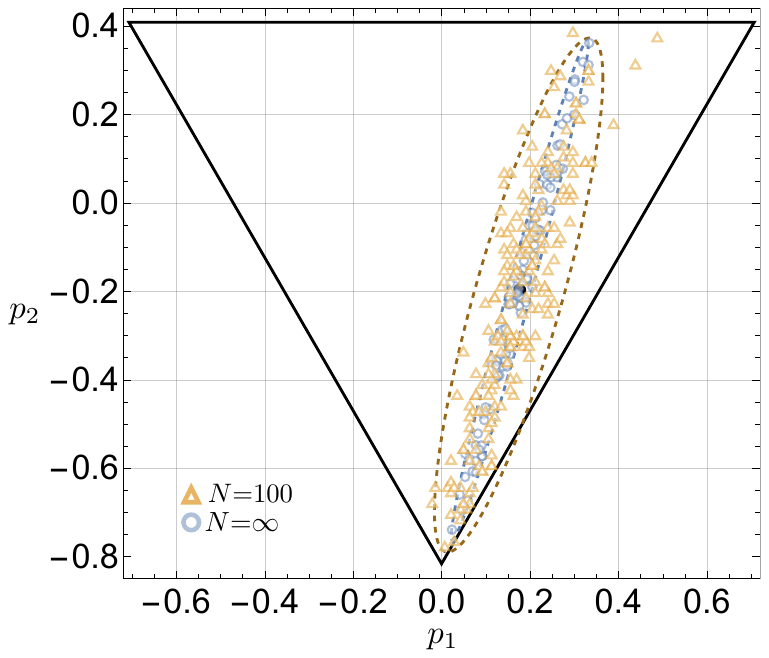}
    \caption{
        \textit{Underestimation of condition number for limited statistics} ---
        We consider random qubit states measured with a random three-outcome POVM with unit-rank operators. Each state is thus represented as a length-3 probability vector. Exploiting the normalisation, each such vector can be projected onto the two dimensions tangent space to the 3-dimensional simplex, $p_1$ and $p_2$ in the figure.
        The blue dots represent here the probability vector associated to each state. The orange triangles, those obtained sampling from the same probability vectors with finite statistics (in this case $N=100$ sampled were used).
        The dashed ellipses are drawn using as principal axes the principal components of the corresponding matrices of probabilities, and have principal lengths corresponding to the associated singular values.
        The condition number of $P_N$ is then proportional to the ratio between largest and smallest principal axes of its ellipse.
        As clearly seen from the spread of the orange triangles here, points estimated from finite statistics result in larger smallest singular values, and therefore smaller condition numbers.
    }
    \label{fig:condition_number_mess}
\end{figure}

%\subsection{Reconstruction efficiency \mauro{spostiamo questa sezione in appendice}}
As in the main text, we denote with $P_N$ the matrix whose columns are the frequencies obtained estimating the outcome probabilities with $N$ samples for different training states. The columns of $P_N$ are thus finite-sample estimates of the columns of $P$.
Even though $\kappa(P_N)$ can be larger for larger $N$, the corresponding estimation error always decreases with $N$, because the inaccuracies in the estimated probability vectors also decrease with $N$ counteracting the increased noise sensitivity flagged by $\kappa(P_N)$.
A rough intuition for why $\kappa(P_N)$ often increases with $N$ can be obtained as follows: an arbitrary matrix $P_N$ can be pictured as the ellipsoid that it maps the unit (hyper)sphere to. The singular values of $P_N$ are then proportional to the lengths of the principal axes of this ellipsoid. The condition number equals the ratio between larger and smaller (\textit{nonzero}) principal axes.
In our case, the columns of $P_N$ are subject to statistical noise that causes them to fluctuate by a quantity of the order of $1/N$.
Consequently, directions corresponding to singular values smaller than the statistical noise, will appear larger, with lengths in the order of $\sim 1/N$.
The overall result is that for small $N$ the directions corresponding to the smallest singular values of $P_N$ might appear larger, thus causing $\kappa(P_N)$ to be underestimated. This underestimation will become negligible when the statistical noise has magnitude significantly smaller than the smallest singular value of $P_N$. This also explains why the underestimation is most prominent in situations where $\kappa(P_N)$ is larger, which is generally due to the smallest singular value of $P_N$ being smaller.
This phenomenon is displayed in a simple case in~\cref{fig:condition_number_mess}.

% \addLI{Another interesting and tightly related feature is the behaviour of the MSE in the cases of finite and infinite statistics at training and testing stages, shown in~\cref{fig:MSE_vs_numMeas_linear}, which relates with the behavior of the condition number at finite statistics. More specifically, high values of $\kappa(P_N)$ correspond to higher errors in the estimation of the correct weight matrix $W$ for finite training statistics.}

\begin{figure}
    \centering
    \includegraphics[width=0.9\linewidth]{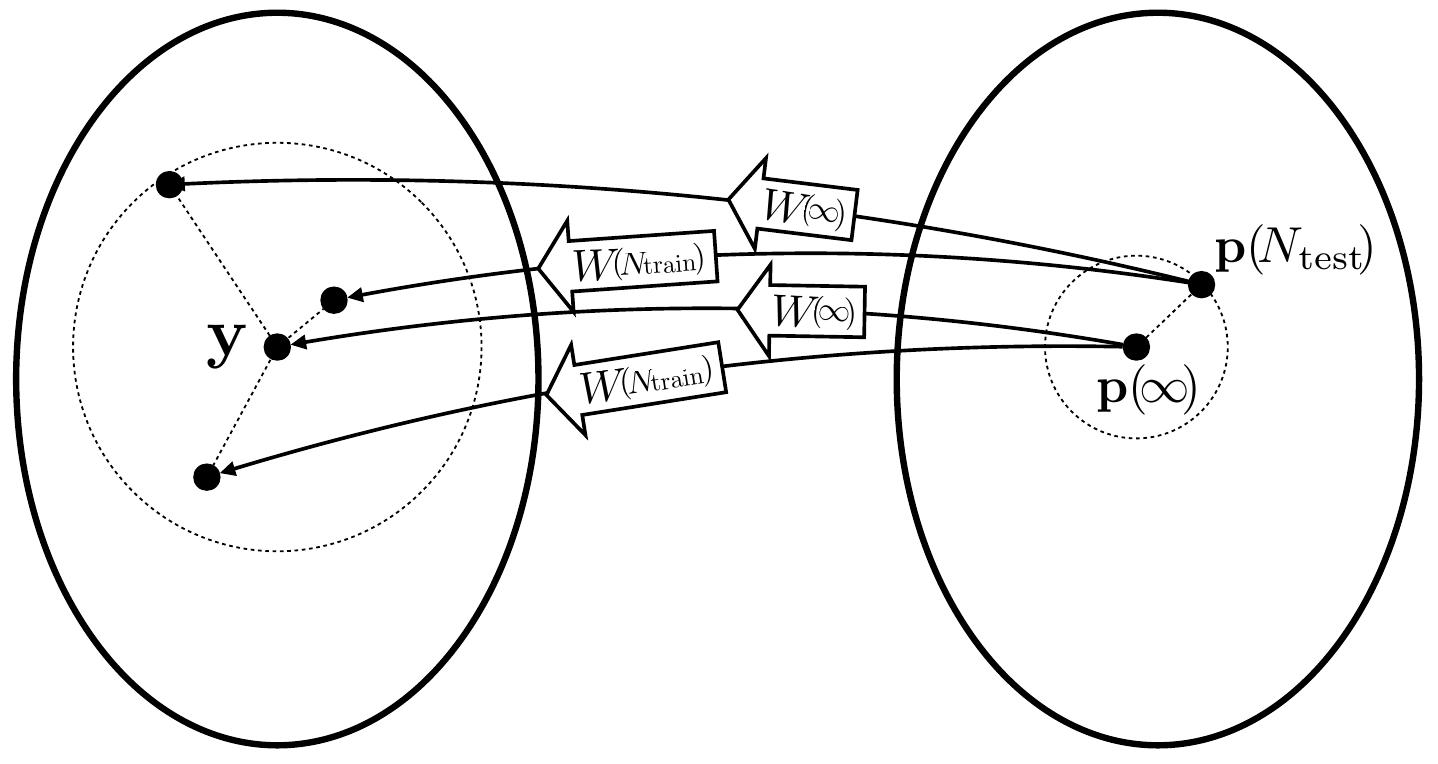}
    \caption{Schematic illustration of how statistical estimation errors are amplified for different training and testing statistics $N_{\rm train}$ and $N_{\rm test}$. When $N_{\rm train}\to\infty$, the associated weight matrix $W(\infty)$ sends ideal probabilities $\mathbf p(\infty)$ to ideally estimated expectation values $\mathbf y$. However, if $N_{\rm test}$ is finite, then $W(\infty)$ operates on estimates probabilities $\mathbf p(N_{\rm test})$, and the statistical errors in $\mathbf p(N_{\rm test})$ might be significantly amplified by $W(\infty)$.
    On the other hand, using finite training statistics, $W(N_{\rm train})$ incorrectly estimates $\mathbf p(\infty)$, but might amplify the errors in $\mathbf p(N_{\rm train})$ less than $W(\infty)$.
    The overall effect is that using $N_{\rm train}\gg N_{\rm test}$ might negatively impact the estimation MSE.
    } \label{fig:graphics_W_differentstatistics}
\end{figure}

\addLI{Another way to understand the potentially odd behaviour of the condition number shown in~\cref{fig:MSE_vs_numMeas_linear}, and in particular its increasing with the number of training statistics $N_{\rm train}$, is to observe that when $N_{\rm train}$ is significantly larger than the testing statistics $N_{\rm test}$, one might incur in a phenomenon analogous to overfitting.
Indeed, even though using large $N_{\rm train}\gg1$ results in a weight matrix $W(\infty)$ which sends ideal output probabilities $\mathbf p(\infty)$ to the corresponding accurate expectation values, it is possible that this $W(\infty)$ significantly amplifies errors in the probabilities $\mathbf p(N_{\rm test})$ estimated with finite statistics, and thus results in an overall larger estimation error, unless $N_{\rm test}$ is also large enough to overcome this effect.
For this reason, having $N_{\rm train}\gg N_{\rm test}$ may result in overall decreased performances, because even though $W(\infty)$ sends ideal probabilities $\mathbf p(\infty)$ 
into perfectly estimated expectation values, noisy probabilities $\mathbf p(N_{\rm test})$ might be sent to estimated expectation values worse than those that would have been produced with $W(N_{\rm train})$.
This phenomenon is shown schematically in~\cref{fig:graphics_W_differentstatistics}.
}

\end{document}